\documentclass[12pt]{article} \usepackage{a4wide,epsf}

\begin{document}
\begin{titlepage}
\begin{flushright}
  CERN-TH/95-325\\
  hep-ph/9512269
\end{flushright}
\vspace{.1cm}
\begin{center}
  {\Large\bf High-order behaviour and summation methods in
    perturbative QCD}\\
  \vspace{1.3cm} {Jan Fischer\footnote{Permanent address: Institute of
      Physics, Academy of Sciences of the Czech Republic, Na Slovance
      2, CZ-180 40 Praha 8, Czech Republic;\,\, e-mail address:
      fischer@fzu.cz, fischerj@cernvm.cern.ch}}\\
  {\em CERN, Theory Division}\\
  {\em CH-1211 Geneva 23, Switzerland}\\
  \vspace{.5cm}

\end{center}
\vspace{3cm}
\begin{center}
  {\bf Abstract}
\end{center}
{\small
  After reviewing basic facts about large-order behaviour of
  perturbation expansions in various fields of physics, I consider
  several alternatives to the Borel summation method and discuss their
  relevance to different physical situations. Then I convey news about
  the singularities in the Borel plane, and discuss the topical
  subject of the resummation of renormalon chains and its application
  in various QCD processes.}

\vspace{1cm}
\begin{center}
  {\it Invited Talk at the Second German-Polish Symposium on\\
    "New Ideas in the Theory of Fundamental Interactions",\\
    September 1995, Zakopane, Poland}
\end{center}

\vspace{2cm}
\begin{flushleft}
  CERN-TH/95-325\\
  December 1995
\end{flushleft}
\end{titlepage}

\newpage

\setcounter{footnote}{0}

\section{Introduction. Some history}

The problem of the high-order behaviour of the perturbation-expansion
coefficients in field theory calculations has aroused new interest
during the past two years. Particular attention has been paid to power
corrections to QCD predictions for hard scattering processes. One can
point out two reasons for this growing interest:
\begin{itemize}
\item The theoretical problem of how physical observables can be
  reconstructed from their (often divergent) power expansions.

\item The pragmatic need to assess the usefulness of performing the
  extensive evaluations of multi-loop Feynman diagrams in QCD. Much
  effort has been devoted to the computation of higher-order QCD
  perturbative corrections; in some cases third-order approximations
  are known, and we now seem to be at the border of what can be
  carried out analytically or numerically in high-order perturbative
  calculations. So, why next order? If the series is an asymptotic
  one, the next order may represent no improvement with respect to the
  lower-order result. On the contrary, at a certain order it will lead
  to deterioration.

\end{itemize}

There are good reasons to believe that the vanishing convergence
radius of perturbation expansions of physically relevant quantities is
a general feature of quantum theory; it has been proved in most of the
theories and models considered. Bender and Wu \cite{Bender-Wu} showed
in 1971 that perturbation theory for one-dimensional quantum mechanics
with a polynomial potential is divergent, and they also discussed the
very-large-order behaviour of the perturbation coefficients. In 1976,
Lipatov \cite{Lipatov} obtained the same results for massless
renormalizable scalar field theories.

Br\'ezin, Le Guillou and Zinn-Justin \cite{Brezin}, applying
independently the same method to anharmonic oscillations in quantum
mechanics, were able to rederive and to generalize the results
obtained by Bender and Wu.

In QED, after the pioneering work of Dyson \cite{Dyson}, a number of
papers appeared discussing the analyticity properties of the Green
functions (some of them are cited in refs. \cite{Stevenson1} and
\cite{LeGuillou}). The growth like $n!$ (where $n$ is the order of
approximation) has two sources:

1. The number of diagrams grows like $n!$, each diagram giving a
contribution of the order of 1.


2. There are types of diagrams for which the amplitude itself grows
like $n!$ \cite{GrN}.


Gross and Periwal \cite{Periwal} proved in 1988 that perturbation
theory for the bosonic string diverges for any value of the coupling
constant and is not Borel summable.

The situation in QCD is particularly complex, not only because the
expansion coefficients behave like $n!$ and are of non-alternating
sign, but also due to the strong dependence of a truncated series on
the renormalization prescription. Of particular interest are
situations in which the kinematic regime allows one quantity, say $M$
(momentum, rest mass), to have a large value. Then the matrix element
of a QCD operator ${\cal O}$ containing quark fields is represented by
means of the operator product expansion in inverse powers of $M$:
\begin{equation}
{\cal O} = C_{1}(M/\mu ) {\cal O}_{1}(\mu) + \frac{1}{M}C_{2}(M/\mu )
{\cal O}_{2}(\mu) + O(1/M^2) , \label{Sachra}
\end{equation}
where $\mu $ is the renormalization scale, ${\cal O}_{i}$ are local
operators of the theory (ordered by their dimension) and
$C_{i}(M/\mu)$ are the expansion coefficients, to be calculated in
perturbation theory. (As ${\cal O}$ is independent of the
renormalization prescription, the renormalization-prescription
dependence of the quantities on the right-hand side must mutually
cancel.)

Relation (\ref{Sachra}) allows for the separation of short- and
long-distance effects in the process. Perturbation theory does not
allow (as will be discussed below) an unambiguous computation of the
coefficients $C_{i}(M/\mu)$. This generates an ambiguity of the order
of $1/M$ (see \cite{Sachrajda}) in the determination of
$C_{1}(M/\mu)$, which implies that one should not include the $O(1/M)$
term in (\ref{Sachra}) until $C_{1}(M/\mu)$ has been correctly
computed.  Leaving aside for a while the question of what a correct
computation of $C_{1}(M/\mu)$ means, we conclude that the problem of a
perturbative determination of $C_{1}(M/\mu)$ has to be solved before
passing to, or simultaneously with, higher terms in the expansion
(\ref{Sachra}). In my talk I shall focus on the former subject,
referring for the latter topic to the recent review talk of ref.
\cite{Sachrajda} and references therein.

\begin{table}
\begin{tabular}{lll}
{\bf Theory}  & {\bf Notation and references} & {\bf High-order
behaviour} \\[10pt]
\phantom{} 1 Disp. relation & $ {\rm Im}f(z) \sim z^{-b} {\rm e}^{-a/z}$ &
$f_n \sim (-a)^n \Gamma (n+b)$  \\
\phantom{} 2 Anharmonic  & $E_m = \sum_{n=0}^{\infty} E_{m,n} g^n$ &
 $E_{m,n} \sim (-\frac{3}{4})^n n!$  \\ \phantom{2 } oscillator
& \cite{Bender-Wu} \\
\phantom{} 3 Anharmonic  & $I(g) \sim \int_{-\infty}^{\infty}
{\rm e}^{(-x^2 /2 - gx^4 /4)} {\rm d}x$ &
$I_n \sim (-1)^n \frac{4^{-n}}{n} n! $ \\ \phantom{3 } oscillator, \,\,
$\phi^4$ & \cite{Bender-Wu,BogomFat1,Parisi} \\
\phantom{} 4 Instantons,  & $m \geq 6$, \cite{Lipatov,BogomFat1} & $C_n (m)
\sim (-a)^n {\rm e}^{n(1-m/2)} n^{(m+D)/2}$  \\
\phantom{4 } $D=\frac{2m}{m-2}$
& $m=4$,  \cite{Lipatov} & $C_n (4) \sim (\frac{n}{16 \pi ^2 e})^n n^4 $ \\
\phantom{} 5 Field theories & $Z=\sum_{n=0}^{\infty} Z_{n}
(-\frac{e^2}{4\pi})^n$ &
 $Z_{n} \sim C n^{b} A^{-n} n!$ \\ \phantom{5}
 without fermions & \cite{Brezin},\cite{Itzykson}-\cite{Hurst}\\
\phantom{} 6 Field theories & \cite{Parisi} & $f_{n} \sim \Gamma (n
\frac{d-2}{d})$ \\
\phantom{6 } with fermions & \cite{BogomFat2} & $f_n
\sim (-\alpha)^n A^{n/2} \Gamma(\frac{1}{2} n)$ \\
\phantom{} 7 Yukawa theories & \cite{Parisi} & $Z_{n} \sim n^{-\alpha}  A^{-n}
{\rm cos}(\frac{2 \pi n}{d}) \Gamma (n \frac{d-2}{d})$ \\
\phantom{Yuk} $d=2$ & \cite{Fry} & $Z_{n} \sim A^{-n}
(\ln\, n)^n $ \\
\phantom{} 8 QED & \cite{BogomFat2,Balian} & $Z_{n} \sim (-1)^{n}A^{n} \Gamma
(\frac{1}{2}n)$ \\
\phantom{} 9 QCD & \cite{Mueller} & $\sim A^n n^{\gamma} n!$\\
10 Bosonic strings  & $h$ is \# of handles,\,\, \cite{Periwal} &
$\sim h!$
\end{tabular}
\caption[]{{\bf High-order behaviour of perturbation expansion coefficients}}
\end{table}

Table 1 gives a survey of the large-order behaviour of the expansion
coefficients in some typical theories and models. It is intended for
first information and should not be used for systematic analyses
because some important conditions or restrictions are not mentioned.
(Let me also point out that references in the table as well as in the
talk as a whole are often made not to the original papers but rather
to reviews or more recent papers.
As a result, a number of relevant valuable papers are not quoted; I
apologize to their authors.)  For details the reader is referred to
the original papers and to the anthology by Le Guillou and Zinn-Justin
\cite{LeGuillou} (which contains a list of references up to \,1990).

The series are divergent in all the cases indicated, with vanishing
radius of convergence. In this connection, two remarks are in order.
The first is that the estimates are based only on certain subclasses
of higher-order diagrams which, in the case of QED (QCD), are obtained
by inserting an arbitrary number of fermion loops into the photon
(gluon) lines of the lowest-order radiative correction.  It is not
known which additional contributions come from other diagrams; whether
they are negligible or give the same or even a greater contribution,
or finally whether they cause cancellations in the original estimate.

The second remark I want to make is as follows. In most of the items
of Table 1, the coefficients exhibit the same, factorial, high-order
growth (items 1 -- 5, 9 and 10).  It would however be misleading to
conclude that all these theories face equal divergence and ambiguity
problems, which could be treated in the same way. Knowledge of the
high-order behaviour shows only one part of the problem of a
perturbation expansion, the other ones being those of summability and
of uniqueness of the summability prescription. Some details are
discussed in sections 2 and 3.

\section{Useful facts on power series}

Certain important facts on power series are overlooked in physical
considerations. It may therefore be useful to recall some of them here
because spontaneous intuition is often misleading.

1. The divergence of a perturbation expansion does not signal an
inconsistency or ambiguity in the theory. (See an analogue in item 5.)

The problem is not that of convergence or divergence, but whether the
expansion uniquely determines the function or not. The method of
Feynman diagrams allows one to find, at least in principle, all
coefficients of the perturbation series, which may determine the
function uniquely even if the series is divergent, and may not do so
even if it is convergent. This depends on additional conditions.

2. The requirement of asymptoticity of a perturbation series,
\begin{equation}
f(z) \sim \sum_{n=0}^{\infty}a_{n}z^{n} , \label{as}
\end{equation}
is not a formal assumption. It has physical content.

When a perturbation series is divergent, it is usually reinterpreted
as an asymptotic series. This is a weaker assumption, but not a
technical one. It is by no means trivial; (\ref{as}) for $z
\rightarrow 0$ physically means that there is a very smooth transition
between the system with interaction and the system without it. For
certain classes of observables the perturbation series is believed to
be an asymptotic expansion.

3. If $f(z)$ is singular at the origin, its asymptotic expansion
(\ref{as}) may still be a convergent series.

Consider for example the function $f(z) = g(z) + A {\rm e}^{-\alpha
  /z}$ with $A$ real and $\alpha$ positive, where $g(z)$ is analytic
at the origin: the asymptotic expansion of the singular function
$f(z)$ in the right complex half-plane is a convergent series.  This
convergent series is asymptotic to many functions (most of which are
singular at the origin), but its values converge only to one of them,
$g(z)$. They do so inside the Taylor circle, which extends to the
nearest singularity of $g(z)$.

4.  A very violent behaviour of the expansion coefficients $a_n$ at $n
\rightarrow \infty$ might make us expect that no function with the
property $f(z) \sim \sum_{n=0}^{\infty}a_{n}z^{n}$ would exist. This
fear is not justified; it was proved by Borel and Carleman (see
\cite{Hardy} for details) that there are analytic functions
corresponding to arbitrary asymptotic power series.

5. The Borel non-summability of a perturbation expansion alone does
not signal an inconsistency or ambiguity in the theory.

The Borel procedure is just one of many possible summation methods and
need not be applicable always and everywhere. The problem is to find a
method (not necessarily the Borel one) which is appropriate for the
case considered.

6. If the $a_n$ behave very violently at $n \rightarrow \infty$ (so
that the Borel series $\sum_{n=0}^{\infty}\frac{a_n}{n!}z^n$ has zero
convergence radius) one might expect that it would be sufficient to
replace $n!$ in the denominator by a sequence ${b_n}$ that grows
faster than $n!$, in order to reach a more efficient suppression of
the $a_{n}$. If this is done, the price to pay usually is that
stronger conditions on analyticity will be required for the summation
procedure to be unambiguous. Analyticity of the expanded function must
be examined simultaneously with the asymptotic expansion, otherwise
the same series can be summed to different functions.

We can conclude (and will elaborate below) that uniqueness of a
summation procedure (in other words, recoverability of a function by
means of its asymptotic series, see \cite{Stevenson1}) requires a
balance between the high-order behaviour of the series and the
analyticity domain of $f(z)$. A violent high-order behaviour can lead
to a unique definition only if "enough analyticity" is available.

\section{Analyticity vs. high-order behaviour: \\
  balance for uniqueness}

How to deal with divergent series and under which conditions a power
series can uniquely determine the expanded function are questions of
fundamental importance in quantum theory. Power expansions are badly
needed in physics, but additional conditions are required to ensure
that they have precise meaning. These additional conditions should
reflect some physical features of the system.

The fact that the expansion coefficients, which grow asymptotically
like $n!$, are of constant, non-alternating sign, is the origin of
most problems connected with the nonuniqueness of perturbative
expansions in QCD.

Let us discuss a simple example to illustrate this crucial point.
Consider a generic quantity $D$, calculated in perturbation theory
with coupling $z$, \begin{equation} D(z) = \sum_{n=0}^{\infty}a_n z^n
  \,.\label{1}
\end{equation}
This can be rewritten as
\begin{equation}
D(z) = \sum_{n=0}^{\infty}a_n z^n (1/n!) \int_{0}^{\infty}{\rm d}t
{\rm e}^{-t}t^n
\,.\label{2}
\end{equation}
If the series (\ref{1}) has a non-vanishing convergence radius $r$,
the integration in (\ref{2}) can be exchanged with the sum. If, on the
other hand, the convergence radius is zero, $r=0$, we can give the
series meaning by exchanging the order of integration and summation.
In either case we obtain
\begin{equation}
D(z) =\int_{0}^{\infty}{\rm d}t{\rm e}^{-t} \sum_{0}^{\infty}a_n
\frac{(zt)^n}{n!}
=\int_{0}^{\infty}{\rm d}t{\rm e}^{-t}B(zt)  \,,\label{3}
\end{equation}
where $B(zt)$ is the Borel transform of $D(z)$. Taking $a_{n}= n!$
(finite-order coefficients are irrelevant for the character of
singularities) we obtain
\begin{equation}
D(z) =\int_{0}^{\infty}{\rm d}t{\rm e}^{-t}\frac{1}{1 - zt}
\,.\label{4}
\end{equation}
This integral does not exist for $z$ positive, nor is the Borel sum of
such a series defined. The summation can be defined in many ways;
there are infinitely many functions with the asymptotic expansion
$\sum_{n=0}^{\infty}n!z^{n}$ .

Note that the non-uniqueness of the summation prescription is not a
mathematical difficulty; it rather signals lack (or insufficient use)
of physics in the theory. There are two kinds of conditions for a
function $f(z)$ to be uniquely determined by its asymptotic expansion
$\sum_{0}^{\infty}a_n z^n$. The function must

i) have a sufficiently large analyticity domain $K$

ii) satisfy upper bounds (uniform in $z$ and $N$) on the remainder
$f(z)-\sum_{0}^{N}a_{n}z^{n}$ for each $N$ above a certain value.
When $K$ has a small opening angle at the origin, the inequalities
must be sufficiently restrictive in order to reach uniqueness. If the
angle is large enough, the condition ii) may be weakened.

Let us sketch how this works in the case of the Borel summation
method. The series (\ref{as}) is called Borel summable if

a) its Borel transform,
\begin{equation}
B(t) = \sum_{n=0)}^{\infty} a_n t^n /n! ,
\label{BT}
\end{equation}
converges inside some circle, $\mid t \mid < \delta \, , \,\, \delta >
0$;

b) $B(t)$ has an analytic continuation to a neighbourhood of the
positive real semi-axis ${\rm Re} \,\,t \geq 0$, and

c) the integral
\begin{equation}
g(z) = \frac{1}{z} \int_{0}^{\infty} {\rm e}^{-t/z} B(t) {\rm d} t  \, ,
\label{BS}
\end{equation}
called the Borel sum, converges for some $z \neq 0$.

Nevanlinna \cite{Nevan} gave the following criterion of Borel
summability:

Let $f(z)$ be analytic in the domain $K(\eta)$ defined by the
inequality ${\rm Re}\frac{1}{z} > \frac{1}{\eta}$ (with $\eta$
positive), a disc of radius $\frac{1}{2}\eta$ bisected by the positive
real semi-axis and tangent to the imaginary axis (see Fig. 1a), and
let $f(z)$ have the asymptotic expansion (\ref{as}). If the remainder
$R_{N}(z)$ after subtracting $N$ terms from $f(z)$,
\begin{equation}
R_{N}(z) = f(z) - \sum_{n=0}^{N-1}a_n z^n ,  \label{R}
\end{equation}
is bounded by the inequality
\begin{equation}
\vert R_{N}(z) | < A \sigma^{N} N!  |z|^{N}  \label{bound}
\end{equation}
uniformly for all $z \, \in \, K(\eta)$ and all $N$ above some value
$N_{0}$, then the sum is determined uniquely and has the form
(\ref{BS}) for all $z \, \in \, K(\eta)$.

For other types of regions, similar theorems hold with modified
regions. Details are exposed, along with references, in the review
\cite{Fischer}; a survey of typical cases is given in Table 2,
together with conditions for a unique determination of $f(z)$ from its
asymptotic expansion.
\begin{table}
\begin{tabular}{llll}

{\bf $f(z)$ at $z=0$}  & {\bf Uniform bound} &  {\bf Transform}
  &  {\bf Summation} \\[10pt]

 &   {\bf on $R_N (z)$}  &  & \\

1 Analytic  &  &  &   $\sum_{n=0}^{\infty}a_n z^n$  \\

 &  &  &    is convergent  \\

2 Singular,  &   $A \sigma ^{N} N! |z|^N $    &
  $B(t) =$   &  $g(z) = $  \\

opening angle $ = \pi $  & in $K(\eta)$ and $N > N_{0} $
 & $ \sum_{n=0}^{\infty} \frac{a_n}{n!} t^n$ & $\frac{1}{z}
\int_{0}^{\infty} {\rm e}^{-t/z} B(t){\rm d} t $ \\ \\

3 Singular,  &  $A \sigma ^{N} (N!)^{\rho} |z|^N$
  &   $B_{\rho}(t) =$  & $g_{\rho}(z) =\frac{1}{\rho}
\int_{0}^{\infty} t^{1/ \rho -1}$  \\

opening angle $ > 0 $  & in $K(\eta, \rho)$, $N > N_{0}$
 & $ \sum_{0}^{\infty} \frac{a_{n}t^{n}}{\Gamma(n\rho + 1)}$
 & $ {\rm exp}(-t^{1/\rho}) B_{\rho}(tz) {\rm d}t$ \\ \\

4 Singular,  & $ A \mu (N) |z|^N$  & $M(t) =$  &
$g_{m}(z)= \int_{0}^{\infty} M(tz) $  \\

opening angle $=0$ & in wedge $W$, $N>N_{0}$ & $
\sum_{n=0}^{\infty} \frac{a_n}{\mu (n)} t^n $  &
$ {\rm exp}(-{\rm e}^{t}) {\rm d} t$  \\ \\
\end{tabular}
\caption[]{{\bf Analyticity vs. high orders: a balance is
needed for uniqueness}}
\end{table}

The function $\mu (n)$ used in Table 2 is defined as follows (see
\cite{MorozCMP} for theorems relevant to the wedge-shaped analyticity
region)
\begin{equation}
\mu (n) = \int_{0}^{\infty} {\rm exp}(-{\rm e}^t) t^n {\rm d}t \, .
\label{mu}
\end{equation}
Note the double exponential function in the integrand, which implies a
$(\ln n)^n$ behaviour of the function $\mu(n)$ at large $n$.

\begin{figure}
  $$
  \epsfxsize=\textwidth \epsffile{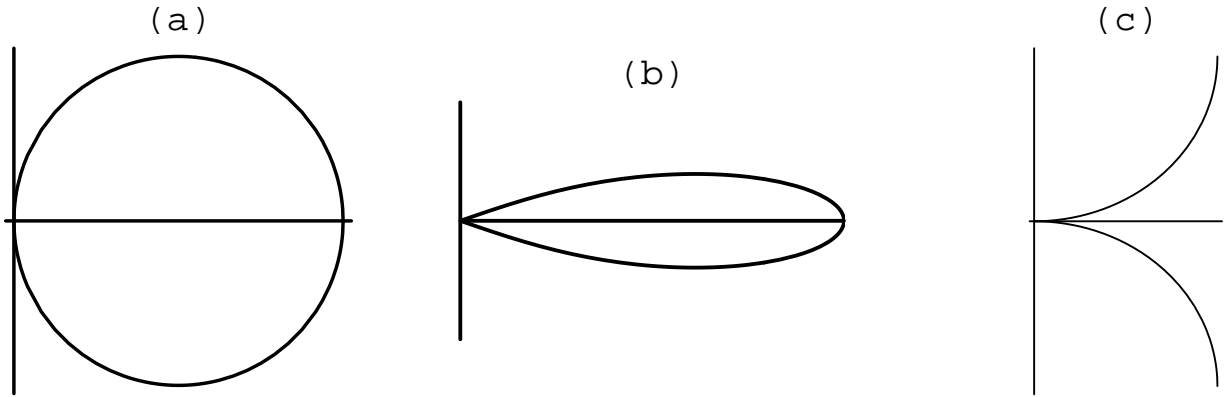}
  $$
\caption[]{Three summation methods (see Table 2) for three
  different analyticity and boundedness domains: (a) the disc
  $K(\eta)$, (b) the drop $K(\eta,\rho)$, and (c) the wedge $W$.
  Crucial is not the size of the domain, but the opening angle at the
  origin.}
\end{figure}

The disc $K(\eta)$, the "drop" $K(\eta, \rho)$ and the wedge $W$ are
depicted in Figs. 1 a, b, and c respectively. The general rule for the
use of Table 2 is as follows: To ascertain uniqueness in the
determination of a function $f(z)$ out of its asymptotic series
(\ref{as}), one has to check if the conditions in all columns in one
row are satisfied for the case considered; otherwise the function is
either overdetermined or not uniquely determined. In QCD, where the
function is, according to \cite{'t Hooft}, analytic only in the wedge
$W$ (see row 4 of Table 2), but the $a_{n}$ behave like $n!$ at large
orders (row 2 of column 2), the function is not uniquely determined by
its perturbation expansion and the situation calls for additional
(non-perturbative) information.

A simultaneous use of Table 1 and Table 2 can tell us to what extent
physical observables and Green's functions can be reconstructed from
the asymptotic series (\ref{as}) in a theory. Taking, for instance,
the QCD large-order behaviour, $A^N N^{\gamma} N!$, from Table 1 (item
9), we obtain in Table 2 the uniform bound in the 2nd row to be valid
on the whole disc $K(\eta)$ of the complex coupling constant plane as
the condition for uniqueness. Of course, we do not expect uniqueness
in perturbative QCD and there is therefore no surprise that QCD
violates the balance required in Table 2; indeed, the actual region of
analyticity of two-point Green functions in QCD is much smaller than
$K(\eta)$, having the form of a horn with zero opening angle at the
origin \footnote{This non-perturbative result (see 't Hooft \cite{'t
    Hooft}) is obtained by combining analyticity and unitarity of
  two-point Green functions in the complex momentum squared plane with
  asymptotic freedom. Of interest is a remark by Moroz \cite{MorozCJP}
  that the use of the Callan-Symanzik equation is not crucial in 't
  Hooft's argument, which works also if theoretical evidence for
  asymptotic freedom is replaced by experimental evidence (if
  available) for a logarithmic dependence of $g^2$ on a sufficiently
  high mass scale.
  }; see row 4 of Table 2 and Fig. 1c.

Table 2 also shows that the large-order behaviour of the coefficients
$a_{n}$ is not the only criterion of the (Borel or some other)
summability of the series (\ref{as}). {\it Even a series with a very
  tame behaviour of the coefficients $a_n$ may not be Borel summable,}
in spite of the radical suppression of the $a_{n}$ by the Borel
factors $n!$. An example is discussed in \cite{MorozCJP}.


\section{Generalized Borel transforms}

1. The functions $B_{\rho}(t)$ and $M(t)$ defined in Table 2 are
generalizations of the Borel transform, which can be used in the
various situations listed in Table 1 to reduce non-uniqueness,
provided some additional information is available. More about the
properties of $B_{\rho}(t)$ and $M(t)$ can be found in
\cite{MorozCMP,MorozCJP,MorozOTH,Fischer} and in references therein.

2. I will now discuss another type of generalization of the notion of
Borel transform, \cite{BrownYaffe,BYZhai,Beneke1}, which makes use of
specific structures of singularities that are typical for QED and QCD.

Let me first make a general remark. Until now I have been mostly
exposing mathematical methods. Now I will pass to some practical
aspects of my talk, assuming that two-point Green's functions have
special singularities, the renormalons, in the Borel plane, a
structure that is now almost universally adopted. These ideas are
based on various mathematical models developed in the late 70's and
early 80's, but nowadays these features are often considered as true
features of Nature. With this reservation I am passing to this
subject.

The electromagnetic current--current correlation function is a useful
example to explain a typical structure of singularities in the complex
coupling-constant plane. Denoting this function $\Pi ^{\mu \nu}$,
\begin{eqnarray}
\Pi ^{\mu \nu} = {\rm i} \int {\rm d}^4 x {\rm e}^{-{\rm i}qx}
\langle 0 | \, {\rm T} (j^{\mu}(x) j^{\nu}(0)) \, | 0 \rangle \\
=(g^{\mu \nu}q^2 - q^{\mu}q^{\nu}) \Pi (-q^2)
\end{eqnarray}
and taking $R$, the ratio of the total cross section for ${\rm e}^+
{\rm e}^- \rightarrow {\rm hadrons}$ to that for ${\rm e}^+ {\rm e}^-
\rightarrow {\rm muon \,\, pairs}$, which is related to its imaginary
part:
\begin{equation}
R(s) = 12 \pi {\rm Im} \, \Pi (s+{\rm i}0^{+}) \, ,
\end{equation}
we introduce a modified quantity ${\tilde {\Pi}}$ defined as
\begin{equation}
{\tilde {\Pi}}(Q^2 ) = - 4\pi ^2 Q^2
\left(\frac{{\rm d}}{{\rm d}Q^2}\right) \Pi (Q^2 )
\end{equation}
(where $Q^{2} = - q^{2}$), to avoid inessential logarithmic terms. The
perturbation expansion of $\Pi$ in the powers of the coupling constant
has the form
\begin{equation}
{\tilde \Pi} \sim 1+\frac{\alpha_{s}(Q^2)}{\pi} \sum_{n=0}^{\infty}
{\tilde \Pi}_{n} (\alpha_{s}(Q^2))^n.
\end{equation}
The dependence of ${\tilde \Pi}$ on $\alpha_{s}(Q^2)$ exhibits a
complex structure of singularities in the coupling constant complex
plane at and around the origin \cite{'t Hooft}. Their nature can be
conveniently displayed when studied in the Borel plane as
singularities of the corresponding Borel transform. Perturbation
theory suggests the following structure of the singularities of the
Borel transform \cite{BYZhai}:

(1) {\em Instanton--anti-instanton pairs} \cite {Lipatov,Brezin}
generate equidistant singularities along the positive real axis
starting at $t = 4$, for $t = 4l, l=1,2,...$. Balitsky \cite{Balitsky}
calculated the behaviour of $R_{{\rm e}^+{\rm e}^- \rightarrow {\rm
    hadrons}}$ near $t=4$ and found the leading ${\rm I - {\bar I}}$
singularity to be a branch point of strength $\frac{11}{6}(N_{\rm
  f}-N)$, where $N$ and $N_{{\rm f}}$ is the number of colours and of
flavours, respectively.

(2) {\em Ultraviolet renormalons} are generated by contributions
behaving as $c_{l+1} \sim (-b_0 /l)^k k!$, for $l=1,2,...$, leading to
singularities located at $t = - 2 l/b_0 $ on the negative real axis.
Near the first of the points, $l=1$, the singularity is $(b_0 t +
2)^{-1+\gamma}$, where $\gamma$ is related to the anomalous dimension
of local operators of dimension 6.

(3) {\em Infrared renormalons} are generated by contributions behaving
as $c_{l+1} \sim (b_0 /l)^k k!$, $l=2,3,...$, leading to singularities
located at $t=l/b_0 $. Near the first of the points, the singularity
behaves \cite{Mueller} as $(b_0 t - 4)^{-1-2\lambda/b_{0}}$.

Brown, Yaffe and Zhai and Beneke \cite{BrownYaffe,BYZhai,Beneke1} use
the information about the structure of the first infrared renormalon.
Expanding ${\rm Im} \, \Pi$ and $\Pi$ in powers of the coupling
constant
with the expansion coefficients $a_{n}$ and $c_{n}$, respectively, and
defining their respective Borel transforms
\begin{equation}
A(z) = \sum_{n=1}^{\infty} \frac{n a_{n}}{\Gamma(n+1)}z^n
\end{equation}
and
\begin{equation}
C(z) = {\tilde c_{0}} + \sum_{n=1}^{\infty} \frac{n c_{n}}
{\Gamma(n+1)}z^n \,,
\end{equation}
they obtain, by comparing the expansion coefficients
\cite{BrownYaffe}, the following relation between $A$ and $C$:
\begin{equation}
A(z) =  \sin (b_0 z) C(z)  \,   ,  \label{AC}
\end{equation}
which turns out to be a consequence of renormalization-group
invariance \cite{Beneke1}. Defining a modified Borel transform ${\cal
  F}(z)$ by
\begin{equation}
{\cal F}(z) = \sum_{n=0}^{\infty} \frac{\Gamma(1+\lambda z)}{
\Gamma(n+1+\lambda z)} f_{n}z^{n}     \label {F}
\end{equation}
(thereby accounting for the first infrared renormalon), the authors of
\cite{BYZhai} and \cite{Beneke1} consider the case of a general beta
function, which they choose in such a scheme that its inverse contains
two terms:
\begin{equation}
1/\beta(g^{2}) = -1/(b_{0}g^{4}) + \lambda /(b_{0}g^2 ) .
\end{equation}
They find that, for this form of $\beta(g^2)$, the relation (\ref{AC})
remains valid also for ${\cal A}(z)$ and ${\cal C}(z)$, the modified
(according to (\ref{F})) Borel transforms of ${\rm Im} \, \Pi$ and
$\Pi$ respectively:
\begin{equation}
{\cal A}(z) =  \sin (b_0 z) \, {\cal C}(z)  \,   .
  \label{ACZhai}
\end{equation}

It was already pointed out that the concept of renormalon is at
present applied to concrete physical situations. This poses a topical
problem: To what extent are renormalons physical concepts and to what
extent are they just artefacts of our, maybe inadequate, formalism.


Table 3 gives a survey of singularities in the Borel plane.

\begin{table}
\begin{tabular}{llll}

& {\bf ${\rm I-{\bar I}}$ pairs} &  {\bf UV renormalons}   & {\bf IR
renormalons} \\  \\

{\bf Position} & $t=4,8,12,...$ & $-2l/b_{0},\,\,\, l=1,2,...$ &  $2l/b_{0}
,\,\,\,l=2,3,...$ \\ \\

{\bf Strength}  &  $\frac{11}{6}(N_{{\rm f}}-N)$    &  $(b_{0}t+1)^{-1+
\gamma}$   &  $(b_{0}t-2)^{-1-2 \lambda /b_0 } $  \\

of the first  & \cite{Balitsky}   &
 & $\lambda = b_1 /b_0 $, \,\, \cite{Mueller}  \\

singularity   &  &
  &  \\

\end{tabular}
\caption[]{{\bf Singularities in the Borel plane for SU($N$) QCD}}
\end{table}

\section{Remarkable phenomena in the Borel plane}

A look at Table 3 reveals that the positions of the singularities in
the Borel plane as well as their strength depend on $N$ and $N_ {{\rm
    f}}$; they will thus change as the latter are varied.  This is
relevant to our discussion because the position and strength of the
singularities nearest to the origin of the Borel plane determine the
large-order properties of the perturbation expansion. This phenomenon
was recently discussed by Lovett-Turner and Maxwell \cite{Maxwell}.
Here I shall repeat a part of the analysis made in ref.
\cite{Maxwell} and summarize a discussion I had with the second author
on this topic.

1. The instanton--anti-instanton pairs are covered by the infrared
renormalons if $N_{\rm f} = N$ (mod 3). This follows from the
condition $4n = 2l/b_{0}$, where $b_{0} = (11N-2N_{{\rm f}})/6$.

2. The first instanton--anti-instanton singularity ($t=4$) disappears
if $N_{{\rm f}} \geq N$ and $N_{{\rm f}} = N$ (mod 6).

3. Take $N=3$ and $N_{{\rm f}} = 15$ as a special case of item 1. Then
$b_{0} = \frac{1}{2}$ and the $l$-th infrared renormalon coincides
with the $l$-th ${\rm I - {\bar I}}$ pair. Since the first renormalon,
$l=1$, does not exist, the leading singularity is the first ${\rm I -
  {\bar I}}$ pair.

4. If $N_{{\rm f}}=16$, $b_{0}=\frac{1}{6}$ and the first infrared
renormalon is located at $t=24$, coinciding with the 6th ${\rm I -
  {\bar I}}$ pair. Items 3 and 4 are examples of situations in which,
contrary to common opinion, instantons play an important role in
large-order behaviour. Analogous situations occur for different values
of $N$; it generally holds that when the number of flavours is
sufficiently high (approaching the flavour-saturation value),
instantons become more important than renormalons.

5. If, on the other hand, the number of flavours decreases, the
importance of renormalons increases. For $N_{{\rm f}}=0$,
$b_{0}=\frac{11}{6} N$ and the density of renormalons is the highest;
for 3 colours, there are 9 infrared renormalons below the first ${\rm
  I - {\bar I}}$ pair, which covers with the 10th of them.

6. The case $N_{{\rm f}}=15$, $N=3$ is very special also from the
point of view of the strength of the infrared renormalon singularity,
whose power is 44 in this case. Because of this, the nearest infrared
singularity disappears, but also the nearest instanton--anti-instanton
singularity disappears, because 15 = 3 (mod 6), as is required in item
2 of this list.

7. It is worth mentioning that if another plane than the Borel one is
chosen (see Table 2, $\rho \neq 1$), all the singularities in the
Borel plane either shrink to the origin or run away towards infinity.

Besides these general properties of the singularities in the Borel
plane, considerable progress has been made in recent years in the
knowledge of the singularities in the Borel plane in special physical
systems and special functions. In particular, much is known in the
case of heavy--light and heavy--heavy quark systems. So, in the
large--$N_{{\rm f}}$ approximation there is a finite set of renormalon
singularities \cite{Beneke2} and a discrete infinity of the renormalon
poles. Also, in the case of heavy--light quark--antiquark systems the
structure of singularities in the Borel plane is known.

The effects of renormalons have been studied also in application to
heavy-quark physics: see \cite{BB2} for renormalons in heavy-quark
pole mass, \cite{NeuSach,BBZ} for renormalons in inclusive heavy-quark
decay rates and \cite{NeuSach} for the case of exclusive decay rates.
The connection between renormalons and power divergences in
heavy-quark physics was studied in \cite{MarSach}.

\section{Resummation of renormalon chains}

The renewed interest in calculating higher-order perturbative
corrections and in examining the high-order behaviour of perturbative
series is intimately related to the investigation of renormalization
scale and scheme dependence of a truncated series as well as to
attempts to estimate its uncalculated remainder. In the past various
criteria for finding a suitable renormalization prescription were
proposed; they are based on an estimate of the size of the remainder.
Examples are Stevenson's principle of minimal sensitivity \cite{PMS},
Grunberg's notion of effective charge \cite{GEF}, and the BLM method
of scale setting \cite{BLM} by Brodsky, Lepage and Mackenzie. The BLM
method, which is based on an analogy with QED, was further developed
recently and I shall briefly comment on it.

The BLM prescription is a method of estimating higher-order
perturbative corrections of a physical quantity, provided that the
first approximation is known. It consists in the use of some "average
virtuality" as scale in the running coupling. Instead of working with
fixed scale,
\begin{equation}
\alpha_{s}(Q^2 ) \int {\rm d}^{4}k F(k, Q)\, ,
\end{equation}
one averages over the logarithm of the gluon momentum:
\begin{equation}
\alpha_{s}(Q_{BLM}^{2} ) \int {\rm d}^{4}k F(k, Q) \equiv
\alpha_{s}(Q^{2}) \int {\rm d}^4 k \left(1-\frac{\beta_{0}}{4\pi}
\alpha_{s}(Q^2)\ln \frac{-k^2}{Q^2}\right) F(k,Q) \,.\,\,\, \label{*}
\end{equation}
This replacement amounts to accounting for higher-order terms in
powers of $\alpha_{s}(Q^2)$ by making use of the renormalization-group
evolution
\begin{equation}
\alpha_{s}(-k^2 )= \alpha_{s}(Q^{2}) \sum_{n=1}^{\infty}
\left(\frac{\beta _{0} \alpha_{s}(Q^{2})}{4\pi }\right)^{n-1}
(-\ln (-k^2/Q^2))^{n-1}
\label{sum}
\end{equation}
and retaining only the first two terms in the sum. This approach was
recently generalized \cite{BenekeBraun,BBB1,Neubert} by introducing
the running coupling constant $\alpha_{s}(k^{2})$ directly into the
vertices of Feynman diagrams, with $k$ being the momentum "flowing"
through the line of the virtual gluon. This modification means
replacement of (\ref{*}) by
\begin{equation}
\alpha_{s}(Q^{*2}) \int {\rm d}^{4}k F(k, Q) =   \int {\rm d}^4 k
\alpha_{s}(-k^2) F(k,Q) \,\,\, \label{**}
\end{equation}
with $\alpha_{s}(x^2)=4\pi/(\beta_{0}\ln
\frac{x^2}{\Lambda^{2}_{QCD}})$.  Note, however, that the beta
function $\beta(\alpha_{s}(Q^{2}))$ is approximated by its first term
only.

This method has been applied in phenomenology to various physical
observables, such as $\tau$ decay hadronic width and heavy-quark pole
mass \cite{BBr,Neubert}, semileptonic B-meson decay \cite{BBB2} and
the Drell--Yan process \cite{BBr}. The method makes maximal use of the
information contained in the one-loop perturbative corrections
combined with the one-loop running of the effective coupling, thereby
providing a natural extension of the BLM scale-fixing prescription.

Ellis et al. \cite{Ellis} use Pad\'e approximants to develop another
method of resumming the QCD perturbative series. The authors test
their method on various known QCD results and find that it works very
well.

\section{Concluding remarks. Criticism}

A typical feature of the present status of the QCD perturbative
corrections is the trend to avoid explicit calculation of higher-order
corrections and, instead, to improve the result by making use of some
of the additional information we may dispose of. Such information may
be, for instance, the renormalization group invariance (which allows
us to introduce the running coupling constant instead of the fixed
one), analyticity, or the structure of singularities in the Borel
plane (where the regular location of singularities, to be approximated
by poles, suggests the use of Pad\'e approximants).

As already mentioned, the notion of renormalons, originally introduced
and used to investigate interesting mathematical models, is now widely
considered to have concrete background in physical phenomena. This
universal belief meets with criticism that argues that the
singularities in the Borel plane are nothing but products of a special
choice of renormalization prescription \cite{Krasnikov}. Methods
generalizing the scale-setting procedure developed by Brodsky, Lepage
and Mackenzie meet with criticism \cite{Chyla} based on the argument
that the approach is not fully independent of the choice of
renormalization prescription. Further research will clarify the issue.

It seems that the present effort in further developing the idea of
renormalon and the corresponding formalism will be helpful in finding
a language appropriate for physical ideas in the non-perturbative
sector. Generalizations of the scale-setting procedures
become valuable by implementing new physical information without
calculating higher-order perturbative corrections (which is not only
cumbersome but also doubtful, due to the divergence of the series), by
using some additional, perturbatively independent information.  This
idea is not new, appearing in theoretical physics whenever technical
difficulties force one to look for methods allowing the exploitation
of all information on the system, including whatever is not adequately
taken into account by the existing formalism.


\vskip 0.6cm

\noindent{\bf Acknowledgements}

I am indebted to Professor Ryszard Raczka and the members of the
Organizing Committee for this marvellous meeting and for creating the
scientific atmosphere. I would like to thank Patricia Ball, Vladimir
Braun, Chris Maxwell and Chris Sachrajda for numerous stimulating
discussions, and Patricia Ball, Alexander Moroz and Matthias Neubert
for carefully reading the manuscript and for valuable comments.

The hospitality of the CERN TH Division and the support of the Grant
Agency of the Czech Republic are gratefully acknowledged.


\vskip 0.6cm


\begin{thebibliography}{article}

\bibitem{Bender-Wu} C.M. Bender and T.T. Wu, Phys.Rev.Lett. {\bf 27}
  (1971) 461; Phys.Rev. {\bf D 7} (1973) 1620; Phys.Rev.Lett. {\bf 37}
  (1976) 117

\bibitem{Lipatov} L.N. Lipatov, Sov.Phys. JETP {\bf 45} (1977) 216;
  JETP Lett. {\bf 25} (1977) 116

\bibitem{Brezin} E. Br\'ezin, J.C. Le Guillou and J. Zinn-Justin,
  Phys.Rev. {\bf D 15} (1977) 1544 and 1558

\bibitem{Dyson} F. J. Dyson, Phys.Rev. {\bf 85} (1952) 631

\bibitem{Stevenson1} P.M. Stevenson, Nucl.Phys. {\bf B 231} (1984) 65
  and references therein

\bibitem {LeGuillou} J.C. Le Guillou and J. Zinn-Justin (Eds.),
  Large-order behaviour of perturbation theory (North-Holland,
  Amsterdam,1990)

\bibitem{GrN} D.J. Gross and A. Neveu, Phys.Rev. {\bf D 10} (1974)
  3235; see also M.C. Berg\`ere and F. David, Phys.Lett. {\bf 135B}
  (1984) 412

\bibitem{Periwal} D.J. Gross and V. Periwal, Phys.Rev.Lett. {\bf 60}
  (1988) 2105

\bibitem{Sachrajda} C.T. Sachrajda, Renormalons, Invited lecture
  presented at the 1995 International Symposium on Lattice Field
  Theory, Melbourne, Australia, July 1995

\bibitem{BogomFat1} E.B. Bogomolny and V.A. Fateev, Phys.Lett. {\bf
    B71} (1977) 93; Phys.Rev. {\bf D 19} (1979) 2974; Phys.Lett. {\bf
    B91} (1980) 431


\bibitem{Parisi} G. Parisi, Phys.Lett. {\bf 76B} (1978) 65

\bibitem{Parisi1} G. Parisi, Phys.Lett. {\bf 66B} (1977) 382

\bibitem{Itzykson} C. Itzykson, G. Parisi and J.-B. Zuber,
  Phys.Rev.Lett. {\bf 38} (1977) 306

\bibitem{Lautrup} B. Lautrup, Phys.Lett. {\bf 69B} (1977) 306

\bibitem{Bukhv} A.P. Bukhvostov and L.N. Lipatov, Phys.Lett. {\bf B
    70} (1977) 48; Sov.Phys. JETP {\bf 46} (1977) 871

\bibitem{Hurst} C.A. Hurst, Proc.Camb.Philos.Soc.{\bf 48} (1952) 625;
  W. Thirring, Helv. Phys. Acta {\bf 26} (1953) 33; A. Peterman, ibid.
  {\bf 26} (1953) 291

\bibitem{BogomFat2} E.B. Bogomolny and V.A. Fateev, Phys.Lett.  {\bf
    B76} (1978) 210

\bibitem{Fry} M. Fry, Phys.Lett. {\bf B 80} (1978) 65; P. Renuard,
  Preprint Ecole Polytechnique (Palaiseau) A 247.1076 (1976)


\bibitem{Balian} R. Balian, C. Itzykson, J.-B. Zuber and G. Parisi,
  Phys.Rev. {\bf D 17} (1978) 1041

\bibitem{Mueller} A. Mueller, Nucl.Phys. {\bf B250} (1985) 327; A.I.
  Vainshtein and V.I. Zakharov, Phys.Rev.Lett. {\bf 73} (1994) 1207

\bibitem{Hardy} G.H. Hardy: Divergent series (Oxford University Press,
  Oxford, 1949)

\bibitem{Nevan} F. Nevanlinna: Zur Theorie der asymptotischen
  Potenzreihen, PhD thesis of the Alexander University, Helsingfors,
  1918; Ann.Acad.Sci.Fennicae Ser. {\bf A12}, No. 3 (1918--19);
  Jahrbuch Fort.Math. {\bf 46} (1916--18) 1463

  A. Sokal, J.Math.Phys. {\bf 21} (1980) 261

\bibitem{Fischer} J. Fischer, Fortschr.Phys. {\bf 42} (1994) 665

\bibitem{'t Hooft} G. 't Hooft: Can we make sense out of "Quantum
  Chromodynamics?", {\it in:} "The Whys in Subnuclear Physics", Proc.
  Erice Summer School, 1977, ed. A. Zichichi (Plenum Press, New York,
  1979), p. 943


\bibitem{MorozCMP} A. Moroz, Commun.Math.Phys. {\bf 133} (1990) 369;
  Quantum field theory as a problem of resummation, PhD. thesis,
  Institute of Physics, Prague, 1991, 'hep-th/9206074'

\bibitem{MorozCJP} A. Moroz, Czech.J.Phys. {\bf 42} (1992) 369

\bibitem{MorozOTH} A. Moroz, Czech.J.Phys. {\bf 40} (1990) 705

\bibitem{BrownYaffe} L.S. Brown and G.L. Yaffe, Phys.Rev. {\bf D 45}
  (1992) R398

\bibitem{BYZhai} L.S. Brown, G.L. Yaffe and C. Zhai, Phys.Rev. {\bf D
    46} (1992) 4712

\bibitem{Beneke1} M. Beneke, Nucl.Phys {\bf B405} (1993) 424; Die
  Struktur der Stoerungsreihe in hohen Ordnungen, PhD. thesis,
  Max-Planck-Institut, Munich, 1992

\bibitem{Balitsky} I.I. Balitsky, Phys.Lett. {\bf B273} (1991) 282

\bibitem{Maxwell} C.N. Lovett-Turner and C.J. Maxwell, Nucl.Phys.
  {\bf B 432} (1994) 147; see also ibid. {\bf B 452} (1995) 213



\bibitem{Beneke2} M. Beneke and V.M. Braun, Phys.Lett. {\bf B 307}
  (1993) 144 and {\bf B 405} (1993) 424; Nucl.Phys. {\bf B 426} (1994)
  30

\bibitem{PMS} P.M. Stevenson, Phys.Rev. {\bf D 23} (1981) 2916

\bibitem{GEF} G. Grunberg, Phys.Rev. {\bf D 29} (1984) 2315

\bibitem{BLM} G.P. Lepage and P.B. Mackenzie, Phys.Rev. {\bf D 48}
  (1993) 2250; S.J. Brodsky, \\G.P. Lepage and P.B. Mackenzie,
  Phys.Rev. {\bf D 28} (1983) 228

\bibitem{BenekeBraun} M. Beneke and V.M. Braun, Phys.Lett. {\bf B 348}
  (1995) 513

\bibitem{NeuSach} M. Neubert and C.T. Sachrajda, Nucl.Phys. {\bf B
    438} (1995) 235

\bibitem{BBZ} M. Beneke, V.M. Braun and V.I. Zakharov, Phys.Rev.Lett.
  {\bf 73} (1994) 3058;\\ M. Luke, A.V. Manohar and M.J.  Savage,
  Phys.Rev. {\bf D 51} (1995) 4924

\bibitem{MarSach} G. Martinelli and C.T. Sachrajda, Phys.Lett. {\bf B
    354} (1995) 423; G. Martinelli, M. Neubert and C.T. Sachrajda,
  preprint CERN-TH 7540/94 (1994)


\bibitem{BBB1} M. Beneke and V.M. Braun, Nucl.Phys. {\bf B 426} (1994)
  301; P. Ball, M. Beneke and V.M. Braun, Nucl.Phys. {\bf B 452}
  (1995) 563

\bibitem{Neubert} M. Neubert, Phys.Rev. {\bf D 51} (1995) 5924

\bibitem{BB2} M. Beneke and V.M. Braun, Nucl.Phys. {\bf B 426} (1994)
  301; I.I. Bigi, M.A. Shifman, N.G. Uraltsev and A.I.  Vainshtein,
  Phys.Rev. {\bf D 50} (1994) 2234

\bibitem{BBB2} P. Ball, M. Beneke and V.M. Braun, Phys.Rev {\bf D 52}
  (1995) 3929

\bibitem{BBr} M. Beneke and V.M. Braun, Nucl.Phys. {\bf B 454} (1995)
  253

\bibitem{Ellis} M.A. Samuel, J. Ellis and M. Karliner, Phys.Rev.Lett.
  {\bf 74} (1995) 4380; J. Ellis,\\ E. Gardi, M. Karliner and M.A.
  Samuel: Pad\'e approximants, Borel transforms and renormalons: the
  Bjorken sum rule as a case study, preprint CERN-TH/95-155,
  hep-ph/9509312

\bibitem{Krasnikov} N.V. Krasnikov and A.A. Pivovarov: Running
  coupling at small momenta, renormalization schemes and renormalons,
  preprint INR 0903/95, hep-ph/9510207. S.V. Faleev and P.G.
  Silvestrov: The status of renormalon, preprint BUDKERINP 95-86,
  hep-ph/9510343

\bibitem{Chyla} J. Ch\'yla, Phys.Lett. {\bf B 356} (1995) 341

\end{thebibliography}
\end{document}